# Polarized electroluminescence from silicon nanostructures

**Nikolay Bagraev[1], Eduard Danilovsky[1], Dmitrii Gets[1], Leonid Klyachkin[1], Andrey Kudryavtsev[1], Roman Kuzmin*[,1], Anna Malyarenko[1], Vladimir Mashkov[2]**

[1] Ioffe Physical-Technical Institute of the Russian Academy of Sciences, 194021, St. Petersburg, Russia
[2] St. Petersburg State Polytechnical University, 195251, St. Petersburg, Russia

* Corresponding author: e-mail roman.kuzmin@mail.ioffe.ru

We present the first findings of the circularly polarized electroluminescence (CPEL) from silicon nanostructures which are the p-type ultra-narrow silicon quantum well (Si-QW) confined by δ-barriers heavily doped with boron. The CPEL dependences on the forward current and lateral electric field show the circularly polarized light emission which appears to be caused by the exciton recombination through the negative-U dipole boron centers at the Si-QW - δ-barriers interface with the assistance of phosphorus donors.

**1. Introduction.** Novel solid-state circularly polarized light-emitting devices are of great interest for micro- and nanoelectronics. Such devices were shown to be prepared, for example, by forming chiral structures on a light-emitting surface [1, 2]. Recently, successfulness of such approach has been documented [3]. However the production of the chiral morphology is outside the scope of standard silicon technology. Therefore nonlinear effects inside nanostructures such as the Bose-Einstein condensation for internal generation of the circularly polarized light are more attractive to be used.

In this work we present the studies of the high degree circularly polarized electroluminescence (CPEL) from the silicon nanostructures, which are the p-type ultra-narrow silicon quantum well, Si-QW, confined by the δ-barriers heavily doped with boron in the concentration of N(B) = $5 \times 10^{21}$ cm$^{-3}$ on the n-type Si (100) surface. The electroluminescence appears to reveal rather high intensity and moreover the high degree of circular polarization even at room temperature. The CPEL mechanism seems to be started from the recombination of the excitons bounded by the negative-U dipole boron centers at the Si-QW - δ-barriers interface with the assistance of phosphorus donors.

**2. Methods.** We started with a 350 μm thick n-Si (100) wafer of 20 $\Omega \times$ cm$^{-1}$ resistivity. The both surfaces of the wafer were previously oxidized in an atmosphere of dry oxygen at the temperature of 1150°C. This preliminary oxidation process was mainly used to accumulate excess fluxes of the self-interstitials and the vacancies in the substrate, which are attained if a thin and a thick surface oxide layer is prepared, respectively. The oxide layer thickness was controlled by varying the oxidation time. After the oxidation, the photolithography was used to open the windows at the operating surface of the wafer to perform a short-time diffusion of boron from the gas phase. This short-time diffusion of boron was carried out at the temperature of 900°C in the presence of dry oxygen and chlorine compounds to achieve a high level of generation of the primary defects in addition to the preliminary accumulated in the substrate. The depth of the ultra-shallow diffusion profile of boron determined by the secondary ion mass spectroscopy (SIMS) method was 8 nm. For further electroluminescence (EL) measurements the gold contacts were prepared on the front surface of the sample using standard photolithography technique, whereas the back side of the substrate was covered by aluminum. The device was designed within frameworks of the Hall geometry with the doping area of size 4.7mm by 0.1mm which is provided by eight 200 μm × 200 μm contacts as shown in Fig. 1a.

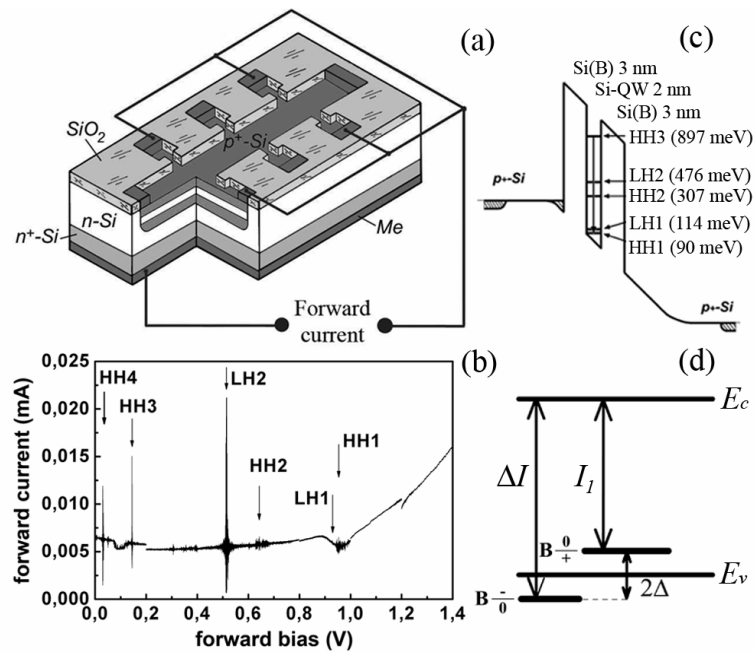

**Figure 1.** (a) A schematic diagram of the device that demonstrates a perspective view of the p-type Si-QW confined by the δ-barriers heavily doped with boron on the n-type Si (100) surface. (b) The forward current-voltage characteristic of the p+-n junction that defines the energies of the two-dimensional subbands of 2D holes in the p-type Si-QW confined by δ-barriers (c). (d) The energies of the ionization of the phone, $I_1$ - (0/+), and correlated, $\Delta I$ - (-/0), electrons that belong to the negative-U boron center. The correlation gap, $2\Delta$, that is able to be transformed in the superconducting gap is equal to 0.044 eV.

The forward current-voltage characteristic of the p+-n junction prepared is shown in Fig. 1b which reveals the system of energy levels inside the Si-QW (Fig. 1c). The samples obtained were set up in a cryostat for the EL measurements in the temperature range from 4.2 to 300 K. The electroluminescence was detected with the MDR-23 monochromator using an InGaAs photodiode and a conventional lock-in technique.

**3. Results and discussions.** The silicon nanostructures obtained were studied by the cyclotron resonance (CR) and ESR measurements as well as using scanning tunneling microscopy (STM) technique. The CR measurements were performed at 3.8 K with a standard Brucker-Physik AG ESR spectrometer at X-band (9.1-9.5 GHz). The CR quenching and the angular dependence with a characteristic 180° symmetry reveal the confining potential inside $p^+$-diffusion profile and its different arrangement in longitudinal and lateral Si-QWs, which are formed naturally between the δ-barriers heavily doped with boron [4, 5]. These Si-QWs were shown to contain the high mobility 2D hole gas that is characterized by long momentum relaxation time of heavy and light holes at 3.8 K [4 – 6].

The δ-barriers heavily doped with boron, $5 \times 10^{21}$ cm$^{-3}$, were found to consist of arrays of the smallest boron doped microdefects with dimensions restricted to 2 nm. The value of the boron concentration determined by the SIMS method and the angular dependences of the ESR spectra showed that each microdefect contains the trigonal dipole boron centers, $B^+ - B^-$, which are caused by the elastic reconstruction along the [111] crystallographic axis of the shallow acceptors as a result of the negative-U reaction: $2B^o \rightarrow B^+ + B^-$ [7]. The energies of the negative-U boron levels were determined using the local tunneling spectroscopy (LTS) technique as well as tunnel current-voltage characteristics. They are $E_c$ + 0.022 eV and $E_c$ – 0.022 eV for (+/0) and (0/-), respectively as represented in Fig. 1d. The trigonal ESR spectrum seems to be evidence of the dynamic magnetic moment that is induced by the exchange interaction between the small hole bipolarons which are formed by the negative-U reconstruction of the boron acceptors [4, 7]. So the dipole boron centers are a basis of nanostructured δ - barriers confining the Si-QW that exhibit the superconductor properties with the critical temperature of $T_c \sim 145$ K defined by the electrical resistivity, thermo-emf and magnetic susceptibility as well as LTS measurements [7, 8]. The presence of the coherent phase of the small hole bipolarons and the orbital moment unfreezing are able to provide the circular polarization of both photo- and electroluminescence.

The EL spectra of the silicon nanostructures detected at the forward current of 20 mA are shown in Fig. 2a for the temperatures of 77 K and 300 K. The main emission line (PB-line) in the low-temperature spectrum is observed at 1.100 eV with the full-width at half-maximum (FWHM) of 18 meV and the phonon replica at 1.037 eV. Under elevated temperature, the line shifts to lower energies following by the band gap temperature dependence giving rise to the intensity drop and to corresponding FWHM broadening up to 63 meV. The photoluminescence (PL) spectra demonstrated the same characteristics (inset in Fig. 2a).

The high degree of the circular polarization in the spectral range of the main emission line is shown in Fig. 2b for both temperature values. It should be noted that the CPEL is not caused by the possible chirality of the nanostructures, because no circular polarization was observed in the studies of the transmission spectra.

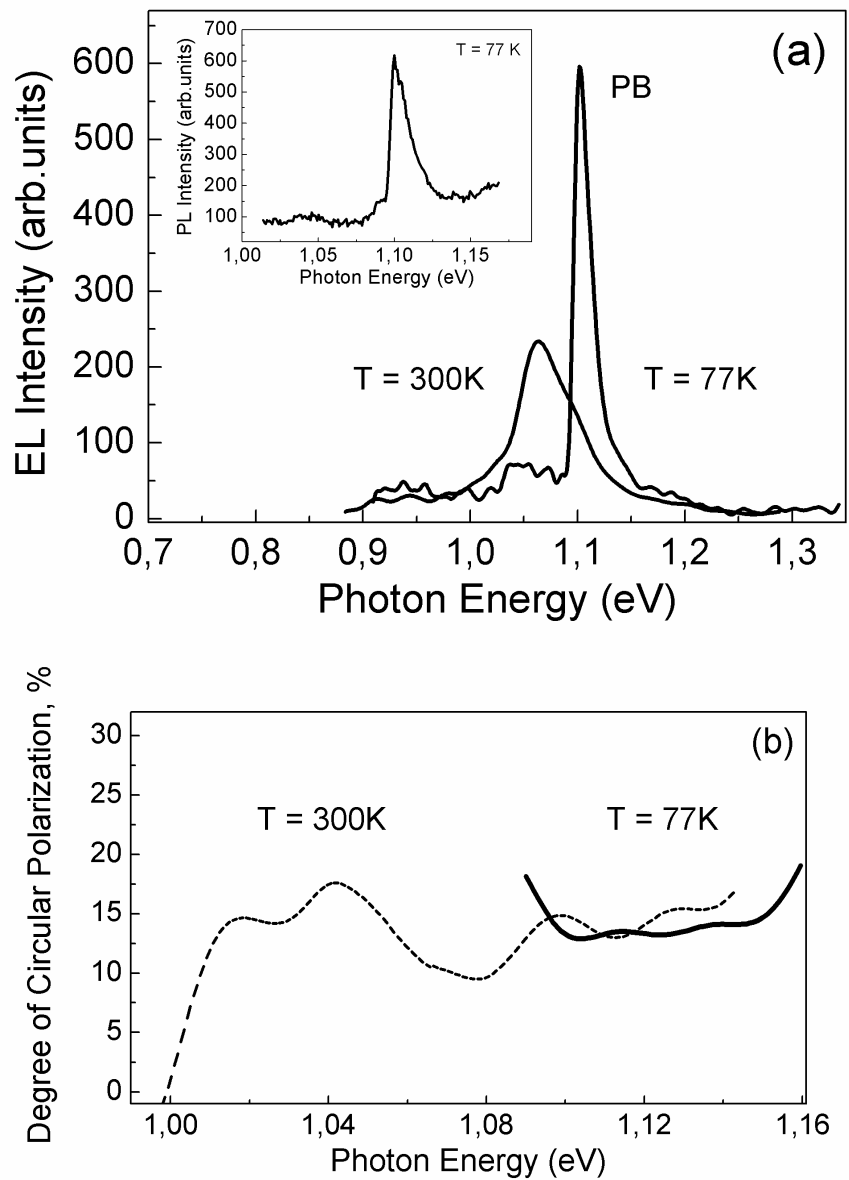

**Figure 2.** (a) The EL spectra detected at forward current of 20 mA for 77K and 300K and the PL spectrum detected for 77 K (see inset). (b) The spectral dependence of the high degree of circular polarization at the temperature of 77 K (solid line) and 300 K (dashed line).

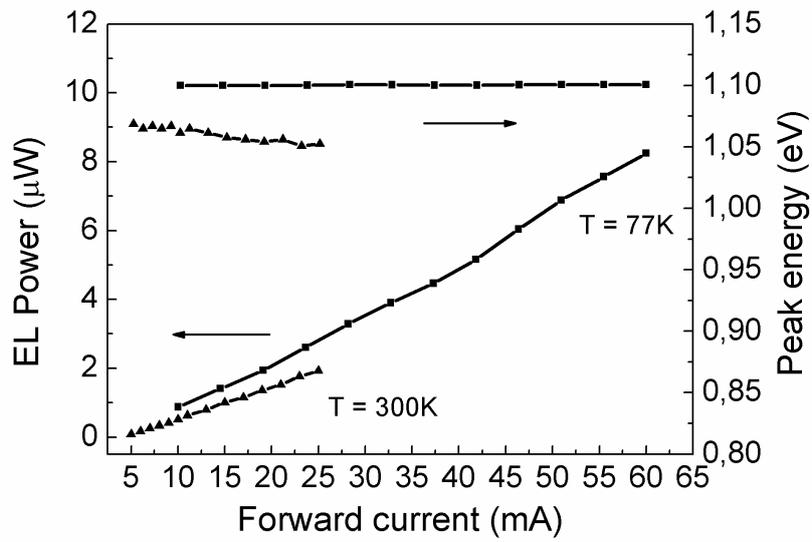

**Figure 3.** The electroluminescence power and the peak energy of the PB-line as a function of forward current at various temperatures: 77 K (squares), 300 K (triangles).

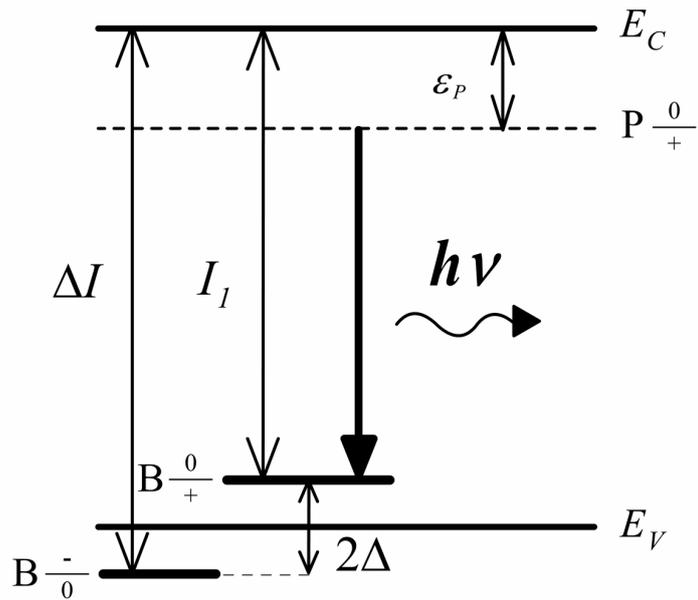

**Figure 4.** The one-electron band diagram of the emission process due to the donor-acceptor recombination, with the electronic and hole states localized on the phosphorus donor and empty, B$^+$, boron center.

The electroluminescence integral intensity of the PB-line is shown in Fig. 3 as a function of forward current, which exhibits the linear character at the temperatures of 77 K and 300 K except low values. Furthermore, the spectral position of the PB-line is independent of the forward current except small shift to lower energies caused by sample overheating at room temperature. The linear character and the independence of spectral position with good accuracy even at very high excitation levels indicate that the valence and conduction band states should not participate in the formation of the PB-line. Taking into account the energy position of the circularly polarized PB-line, the initial and final states of the PB-line seem to be closely related to the phosphorus donors and boron negative-U dipole acceptor centers. The electroluminescence appears to originate at the Si-QW - δ-barriers interface as a result of an emission process due to the donor-acceptor recombination, with the electronic and hole states localized on the phosphorus donor and empty, $B^+$, boron center (Fig. 4).

The contribution of the boron dipoles to the formation of the PB-line is revealed also by the temperature dependence of the electroluminescence intensity, which as can be seen in Fig. 5 increases with the temperature reaching a maximum in a vicinity of 150 K and then decreases rapidly. This EL temperature dependence seems to result from the collapse of the negative-U properties for the dipole boron centers at elevated temperatures. The temperature of the EL intensity quenching is in good agreement with the value of the critical temperature of the superconductor transition, $T_c \sim 145$ K, (see inset in Fig. 5).

The devices designed within frameworks of the Hall geometry allow the studies of the electroluminescence in the presence of additional electric field created by the voltage applied along the surface of the structure. The PB-line integral intensity as a function of this lateral voltage demonstrates the quenching (Fig.6), which is more rapid at high temperatures. Such behavior can be accounted for in terms of two-electron/hole adiabatic potentials for the negative-U boron related center in p-type silicon (Fig. 7). The emission process appears to pass in several stages. At the first stage an electron injected from n-area is captured by a phosphorus ion. The annihilation of the electron and one of the holes on the positive boron ion of the dipole center takes place with the emission of photon. Nevertheless indirect bandgap, the recombination does not require the phonon assistance because of the relaxation moment on the impurity atoms. The $B^0$ state created in the recombination process transits to the $B^-$ state, with the correlated electron captured subsequently (Fig. 7). A great cross-section of this transition caused by negative-U properties of the center results in permanent depletion of the $B^0$-state thereby providing the high intensity of the electroluminescence. Simultaneous reaction $B^- \rightarrow B^0 \rightarrow B^+$ appears to take place on the other boron ion in the presence of two holes injected (Fig. 7). By applying the lateral electric field, it is possible to cause the reciprocal displacement of the $B^-$ - and $B^+$ - states along the [111] and [100] crystallographic axis, respectively.

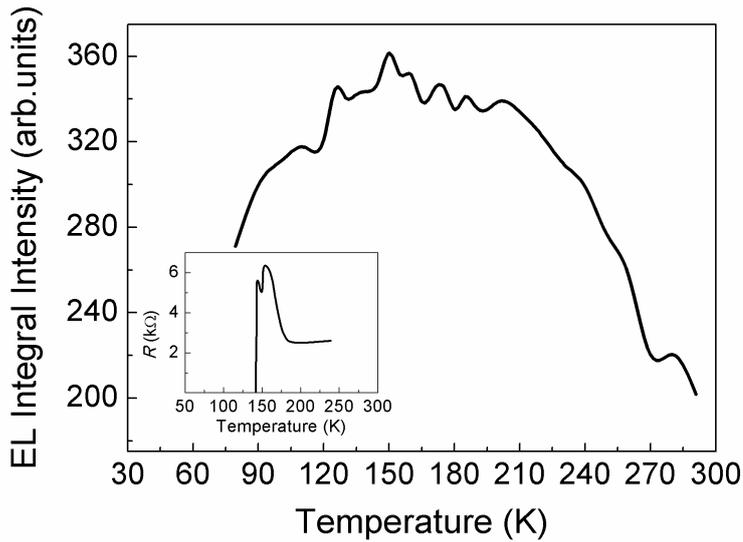
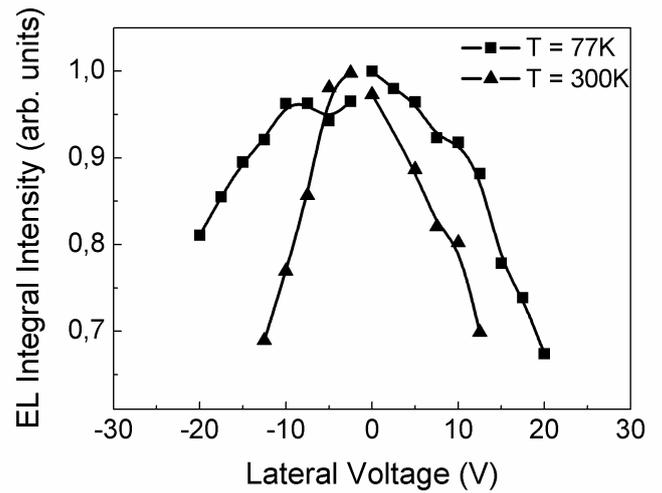

**Figure 5.** The temperature dependence of the integral electroluminescence intensity of the PB-line revealing the maximum in a vicinity of 150 K which correlates with the value of the critical temperature of the superconductor transition, $T_c \sim 145K$, defined by the electrical resistivity (see inset), thermo-emf, magnetic susceptibility and LTS measurements.

**Figure 6.** Integral intensity of the electroluminescence as a function of lateral voltage for the temperatures of 77K and 300K.

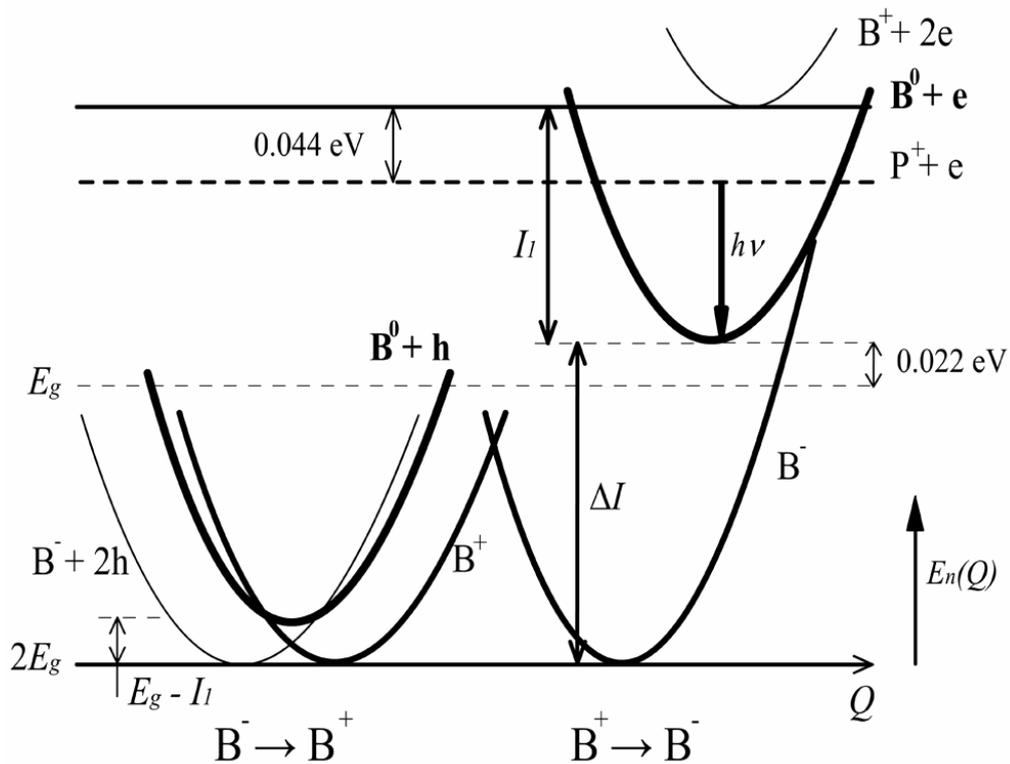

**Figure 7.** The model proposed to the CPEL mechanism based on the exciton recombination through the negative-U dipole boron centers at the Si-QW - δ-barriers interface with the assistance of phosphorus donors.

These shifts due to non-monotonic dependence of the electron-vibration interaction constant on the number of electrons/holes at the center are able to reduce the energy barriers between different boron states thus giving rise to the changes of the population depletion and corresponding quenching in the EL intensity.

**4. Conclusion.** The high degree circularly polarized electroluminescence from the p-type ultra-narrow Si-QW confined by the δ-barriers heavily doped with boron, $5 \times 10^{21}$ cm$^{-3}$, on the n-type Si (100) surface has been demonstrated. The EL studies on different excitation levels and lateral voltages have shown that the mechanism of the efficient circularly polarized electroluminescence seems to be started from the exciton recombination through the negative-U dipole boron centers at the Si-QW - δ-barriers interface with the assistance of phosphorus donors.

**Acknowledgements.** The work was supported by the programme of fundamental studies of the Presidium of the Russian Academy of Sciences "Quantum Physics of Condensed Matter" (grant 9.12); programme of the Swiss National Science Foundation (grant IZ73Z0_127945/1); the Federal Targeted Programme on Research and Development in Priority Areas for the Russian Science and Technology Complex in 2007–2012 (contract no. 02.514.11.4074), the SEVENTH FRAMEWORK PROGRAMME Marie Curie Actions PIRSES-GA-2009-246784 project SPINMET.